\documentclass[aps,pra,amsmath,amssymb,nofootinbib,floatfix]{revtex4-2}

\usepackage[T1]{fontenc}
\usepackage[T1]{tipa}
\usepackage{graphics}
\usepackage{physics}
\usepackage{bm}
\usepackage{times}
\usepackage{natbib}
\usepackage{hyperref}
\usepackage{jabbrv}

\begin{document}

\title{Parallel Quantum-Enhanced Sensing}

\author{Mohammadjavad Dowran$^{1, 2}$, Aye L. Win$^{1}$, Umang Jain$^{1, 2}$, Ashok Kumar$^{1, 3}$, Benjamin J. Lawrie$^{4}$, Raphael C. Pooser$^{5}$, and Alberto M. Marino$^{1, 2, 5,}$\footnote{marino@ou.edu, marinoa@ornl.gov}}

\affiliation{$^{1}$Homer L. Dodge Department of Physics and Astronomy, University of Oklahoma, Norman, OK, 73019, USA\\
$^{2}$Center for Quantum Research and Technology (CQRT), University of Oklahoma, Norman, OK, 73019, USA\\
$^{3}$Indian Institute of Space Science and Technology, Thiruvananthapuram, 695547, India \\
$^{4}$Materials Science and Technology Division, Oak Ridge National Laboratory, Oak Ridge, TN, 37831, USA\\
$^{5}$Quantum Information Science Section, Computational Sciences and Engineering Division, Oak Ridge National Laboratory, Oak Ridge, TN, 37831, USA\footnote{This manuscript has been authored in part by UT-Battelle, LLC, under contract DE-AC05-00OR22725 with the US Department of Energy (DOE). The publisher acknowledges the US government license to provide public access under the DOE Public Access Plan (http://energy.gov/downloads/doe-public-access-plan)}}

\begin{abstract}
Quantum metrology takes advantage of quantum correlations to enhance the sensitivity of sensors and measurement techniques beyond their fundamental classical limit given by the shot noise limit. The use of both temporal and spatial correlations present in quantum states of light can extend quantum-enhanced sensing to a parallel configuration that can simultaneously probe an array of sensors or independently measure multiple parameters. To this end, we use multi-spatial mode twin beams of light, which are characterized by independent quantum-correlated spatial subregions in addition to quantum temporal correlations, to probe a four-sensor quadrant plasmonic array. We show that it is possible to independently and simultaneously measure local changes in refractive index for all four sensors with a quantum enhancement in sensitivity in the range of $22\%$ to $24\%$ over the corresponding classical configuration. These results provide a first step towards highly parallel spatially resolved quantum-enhanced sensing techniques and pave the way toward more complex quantum sensing and quantum imaging platforms.
\end{abstract}

\maketitle

\section{\label{intro}Introduction}
Quantum metrology based on optical fields utilizes quantum-correlated states of light to enhance the sensitivity of measurements and sensing devices beyond the shot noise limit (SNL)~\cite{giovannetti2004quantum,lawrie2019quantum}, as envisioned by the second quantum revolution~\cite{dowling2003quantum,deutsch2020harnessing}. Since the initial proposals for quantum metrology~\cite{caves1981quantum,slusher1985observation}, several applications that harness temporal quantum correlations in quantum states of light have emerged to enhance the sensitivity of devices such as interferometers, magnetometers, plasmonic sensors, accelerometers, spectroscopic measurements, etc.~\cite{pirandola2018advances,lawrie2019quantum,acsphotonics.5b00501,RN1331}. Moreover, spatial quantum correlations are at the heart of the field of quantum imaging~\cite{lugiato2002quantum,genovese2016real,magana2019quantum}, have led to the generation of entangled images~\cite{boyer2008entangled}, and have found their way into applications such as sub-shot noise quantum imaging~\cite{brida2010experimental}, enhanced biological imaging and molecule tracking~\cite{taylor2013biological,taylor2014subdiffraction}, and enhanced measurements of beam displacements~\cite{treps2003quantum}.

Here, we take advantage of  quantum correlations in both the temporal and spatial degrees of freedom of quantum states of light to enable a quantum-enhanced parallel sensing configuration. Such a spatially resolved quantum sensing approach makes it possible to better take advantage of the available quantum resources to perform faster and more efficient measurements by simultaneously estimating multiple parameters or the same parameter multiple times~\cite{spagnolo2012quantum,vidrighin2014joint}. Furthermore, applications such as microscopy, spectroscopy, gravitational field detection, and particle tracking~\cite{lassen2007tools,taylor2013biological,baumgratz2016quantum, muraviev2018massively,im2009plasmonic,acimovic2014lspr,thijssen2014parallel,taylor2016quantum}  can take advantage of such a spatially resolved optical readout for imaging or parallel multi-parameter estimation~\cite{altenburg2018multi,proctor2018multiparameter}.

Two-mode squeezed states, or twin beams, are ideal candidates for parallel quantum-enhanced sensing, as they have been shown to contain quantum correlations in both the temporal and spatial degrees of freedom~\cite{boyer2008entangled,ashok2021IOP}.  These states are generated with a nonlinear process, such as four-wave-mixing (FWM) in alkali atoms~\cite{slusher1985observation}, that generates photons in pairs. In this process, energy conservation leads to temporal quantum correlations between the twin beams that are characterized by noise levels below the SNL, or squeezing, that make it possible to obtain a quantum-based sensitivity enhancement~\cite{lawrie2019quantum}. Additionally, momentum conservation leads to spatial quantum correlations in multi-spatial mode twin beams that are characterized by independently correlated transverse spatial sub-regions. These independently correlated sub-regions can effectively be treated as independent twin beams with the same degree of temporal correlations as the whole beams~\cite{boyer2008generation,kumar2019spatial,ashok2021IOP}.  As a result, each sub-region can be used to probe different spatial regions of a sensor to independently estimate multiple local parameters with a sensitivity beyond the SNL.

We implement a parallel quantum-enhanced sensing configuration by extending our previous work on quantum-enhanced plasmonic sensing~\cite{dowran2018quantum}. In particular, we design and fabricate a quadrant plasmonic structure that consists of four plasmonic sensors that are probed by multi-spatial mode twin beams. As our previous work has shown, plasmonic sensors offer an ideal platform for parallel quantum-enhanced sensing as, in addition to being compatible with quantum states of light to enable the detection of changes in the local refractive index beyond the SNL, they can preserve the spatial information of the quantum light~\cite{PhysRevLett.110.156802,holtfrerich2016toward}. Furthermore, plasmonic sensors~\cite{homola2003present,stewart2008nanostructured,mejia2018plasmonic} have a broad range of applications in biochemical and medical detection as label-free passive sensing devices, and parallel plasmonic sensing configurations based on classical states of light~\cite{im2009plasmonic,acimovic2014lspr,thijssen2014parallel} are already in use. Thus, our work provides a path forward for quantum-enhanced parallel sensing to find its way into practical applications. It also serves as a proof of principle into the possibility of implementing quantum information protocols in parallel~\cite{zoller2005quantum}, as well as a step toward a spatially resolved quantum sensing network, as required for applications such as dark matter detection~\cite{PhysRevD.102.072003,thewindchimecollaboration2022snowmass}.

\section{Experimental Setup}\label{Experiment}
\begin{figure}
  \centering
  \includegraphics{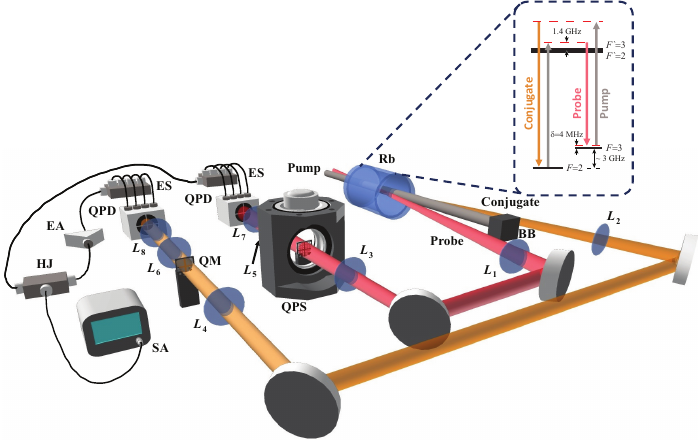}
  \caption{Experimental implementation of parallel quantum-enhanced plasmonic sensing. A FWM process based on the double-$\Lambda$ energy level structure, see inset, in the D1 line of $^{85}$Rb is used to generate multi-spatial mode twin beams. 4$f$ optical systems, composed of $L_{1}$ and $L_{3}$ for the probe and  $L_{2}$ and $L_{4}$ for the conjugate, are used to image the center of the Rb cell (near field) onto the quadrant sensor array and mask (sensing plane). After the plasmonic array and mask, additional 4$f$ optical systems, composed of $L_{5}$ and $L_{7}$ for the probe and $L_{6}$ and $L_{8}$ for the conjugate, are used to image the twin beams onto two quadrant photodetectors to perform intensity-difference measurements between all 16 possible combinations. A variable electronic attenuation ($g$) is used to optimize the quantum noise reduction. QPS: quadrant plasmonic sensor, QM: quadrant mask, QPD: quadrant photodetector, ES: electronic switch, EA: electronic attenuator, HJ: hybrid junction, SA: spectrum analyzer, BB: beam block.}
  \label{Fig:setup}
\end{figure}

To implement a parallel quantum-enhanced sensing configuration that takes advantage of both the temporal and spatial quantum correlations in quantum states of light, we interface multi-spatial mode entangled twin beams generated in an atomic vapor cell with an array of four plasmonic sensors in a quadrant configuration, as shown in the experimental setup in Fig.~\ref{Fig:setup}.  The array of plasmonic sensors is probed by one of the twin beams to detect local modulations in the refractive index of air at each sensor, while the other beam serves as a reference. The reference beam is sent through a quadrant mask to select the spatial sub-regions of the reference beam that are quantum-correlated with the corresponding sub-regions of the probe beam determined by the array of plasmonic sensors.  We image the probe and conjugate beams at the vapor cell center (near field) onto the plane given by the sensor array and mask, or sensing plane, as it has been shown that this makes it possible to maintain the intensity-difference squeezing at low frequencies when selecting sub-regions of a multi-spatial mode twin beams~\cite{wu2021two}.  This is not the case for the far-field (Fourier plane), for which the squeezing can significantly degrade (even leading to noise levels above the SNL) at low frequencies due to two-beam coupling. Finally, the probe and reference beams are imaged onto a balanced differential quadrant detection system to perform independent measurements for each sensor.

\begin{figure}
  \centering
  \includegraphics{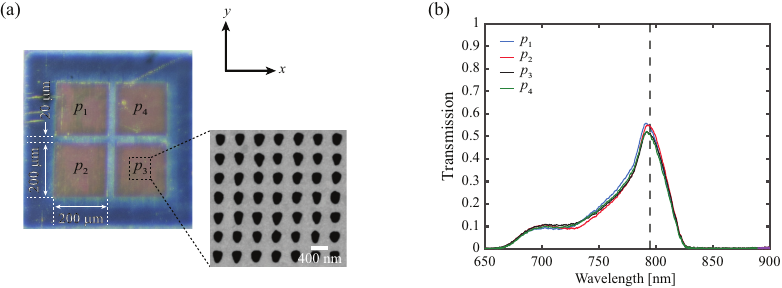}
  \caption{Quadrant plasmonic structure. (a) Microscope image of the quadrant plasmonic structure. The structure consists of four $200~\mu$m$~\times~200~\mu$m nanohole arrays separated by $20~\mu$m. The inset shows an SEM image of the periodic asymmetric nanohole structure used for each of the sensors. The polarization of the incident light is along the $x$ axis. (b) EOT resonance spectra of the quadrant plasmonic sensors. The structure is tilted by 26$^\circ$ about the $y$-axis, normal to the propagation direction of probe beam, to shift the resonances to the probing wavelength of 795~nm, indicated by the vertical dashed line.}
  \label{Fig:PlasmonicSensor}
\end{figure}

The quadrant plasmonic structure is fabricated on a thin silver film ($\approx100$~nm thick) deposited on a glass substrate and consists of four plasmonic sensors, labeled as $p_i$ (with  $i\in\{1,2,3,4\}$) in Fig.~\ref{Fig:PlasmonicSensor}(a), of dimension $200~\mu$m~$\times~200~\mu$m. To isolate the response of the sensors and avoid cross-talk, a gap of $20~\mu$m is left between them on the silver film. Each plasmonic sensor consists of a periodic array of subwavelength asymmetric holes (base of $\approx200$~nm and height $\approx250$~nm) with a pitch of 400~nm in both the $x$ and $y$ directions, as can be seen from the scanning electron microscopy (SEM) image in the inset of Fig.~\ref{Fig:PlasmonicSensor}(a).  Such a nanohole array leads to extraordinary optical transmission (EOT)~\cite{ebbesen1998extraordinary} when plasmons are excited by a resonant electromagnetic field, which allows for large optical transmission through the structures. For our specific design, the EOT response of the four plasmonic sensors is shown in Fig.~\ref{Fig:PlasmonicSensor}(b). To obtain the optimal response from the sensors, the angle of incidence of the light on the structure is scanned to shift their spectra and optimize the transmission and slope at our operational wavelength of 795~nm.  At an optimal angle of $\approx26^{\circ}$, we measure a transmission in the range of $50\%$ to $55\%$ for all sensors.  The quadrant mask used in the path of the reference beam has the same layout as the quadrant plasmonic structure, except that instead of the nanohole arrays, the silver is completely etched out to allow for a clear aperture. We label the windows of the mask as $c_j$ with $j\in\{1,2,3,4\}$, following the same layout as the one for the plasmonic array shown in Fig.~\ref{Fig:PlasmonicSensor}(a). The square $200~\mu$m~$\times200~\mu$m ``windows'' in the mask are measured to have a transmission of $\approx90\%$.

We generate the multi-spatial mode twin beams used to probe the quadrant plasmonic structure with a seeded FWM process~\cite{mccormick2008strong} in a 12~mm long $^{85}$Rb vapor cell held at a temperature of $111^{\circ}$C. As shown in the inset of Fig.~\ref{Fig:setup}, the FWM  is based on a double-$\Lambda$ configuration in which two pump photons are absorbed to generate probe and conjugate photons~\cite{hemmer1995efficient}. In implementing the FWM processes for parallel quantum-enhanced sensing, care needs to be taken to optimize both the temporal and spatial quantum correlations.  In particular, the size of the independently correlated sub-regions of the twin beams, known as the coherence areas~\cite{boyer2008generation}, need to be significantly smaller than the area of each sensor. In general, the size of the coherence area depends on the number of spatial modes supported by the FWM process and can be reduced by increasing the size of the pump beam~\cite{holtfrerich2016control}. Thus, the size of the pump and seed probe beams needs to be optimized to have enough coherence areas in the bright region of the twin beams to perform independent measurements while maintaining a significant level of intensity-difference squeezing.

The FWM is implemented in the D1 line of $^{85}$Rb with a Ti:sapphire laser tuned to 795~nm to generate the required pump and probe beams. The pump beam, which is derived directly from the laser, has a power of $2.2$~W and a Gaussian spatial profile with a $1/e^2$ diameter waist of 3~mm at the center of the vapor cell. The seed probe  is obtained by picking off a low power beam from the laser and double passing it through an acousto-optical modulator (AOM) to red-shift its frequency with respect to the pump beam by $\approx3$~GHz, which corresponds to the hyperfine splitting of the ground state (5$^2$S$_{1/2}$) of $^{85}$Rb. The seed probe, with a Gaussian spatial profile and a $2.5$~mm $1/e^2$ diameter waist at the center of the vapor cell, intersects with the pump beam at an angle of $0.5^{\circ}$. With these optimal beam sizes (see below), we optimize the frequencies of the pump and probe to obtain the largest possible level of squeezing, which leads to  single- and two-photon detunings (see inset in Fig.~\ref{Fig:setup}) of $\Delta=1.4$~GHz and $\delta=4$~MHz, respectively, and $-5.16$~dB of intensity-difference squeezing. It is worth noting that, as shown in our previous work~\cite{dowran2018quantum}, the FWM process is capable of producing twin beams with $-9$~dB of intensity-difference squeezing; however, the parameters required to do so lead to twin beams that do not contain enough coherence areas. This points to a trade-off between producing highly temporally correlated versus multi-spatial-mode twin beams with the FWM process due to the power and size requirements of the pump beam.

To determine optimal sizes for the pump and probe given the amount of power available, we performed spatially dependent intensity-difference measurements of corresponding sub-regions of the probe and conjugate beams. In particular, given the configuration of the plasmonic sensors and mask used in the experiment, we selected different quadrants of the Gaussian-shaped probe and conjugate  with razor blades at right angles to allow the transmission of only a single quadrant of the beam. The razor blades are placed at the sensing plane, and the center of the cell is imaged onto this plane. Given that in the near field corresponding quantum-correlated sub-regions are located at the same position of the twin beams with respect to their center, the razor blades are set to select the same quadrants for the probe and the conjugate beams. For multi-spatial mode twin beams, the level of intensity-difference squeezing is only slightly reduced after cutting when sub-regions significantly larger than the coherence area are selected~\cite{martinelli2003experimental,boyer2008generation,holtfrerich2016control}.  Thus, for our current application of parallel sensing, it is necessary for the coherence area to be significantly smaller than a quadrant of the beam.  We thus optimize the size of the pump and probe such that the squeezing level is only slightly reduced after cutting with the razor blades. The levels of squeezing after cutting with the razor blade are measured by focusing the probe and conjugate beams onto a single quadrant of their corresponding quadrant photodiode and then performing intensity-difference measurements.  Furthermore, we verify that if quadrants that are not expected to be correlated are measured, the noise adds up in quadrature (see Appendix), which points to those quadrants being uncorrelated as needed to perform independent measurements with the different spatial sub-regions of the twin beams.

With the optimal parameters for the FWM process determined, the optical system used to image the center of the cell to the sensing plane is designed to minimize  losses due to the quadrant configuration of the sensor array and mask. The optical systems after the FWM, consisting of 4$f$ imaging systems composed of lenses  $L_{1}$ and $L_{3}$ ($L_{2}$ and $L_{4}$) for the probe (conjugate) with lenses of focal lengths 500~mm and 100~mm, produce probe and conjugate Gaussian beams with $1/e^2$ waist diameters of $360~\mu$m at the sensing plane.  This leads to close to optimal illumination of the sensor array and mask (see Appendix). Note that the measured waist diameters are smaller than expected from the magnification of the optical system.  This is a result of a cross-Kerr effect on the probe beam during the FWM process due to the required strong pump beam. The four sub-regions of the probe beam after transduction through the quadrant plasmonic sensors and the four sub-regions of the conjugate beam after transmission through the mask are then imaged onto two separate quadrant photodetectors (Hamamatsu model S5980/-10 with an active area of $2.48~$mm~$\times~2.48~$mm) with $\approx95\%$ quantum efficiency  at $795$~nm. This is achieved with an  optical system that magnifies the probe and the conjugate such that each quadrant of the probe and conjugate is only incident on one of the quadrants of the photodetector. This optical system consists of an aspheric lens (Thorlabs A240TM-B, effective focal length  EFL$=0.8$~mm, numerical aperture NA$=0.5$), labeled as  $L_{5}$ for the probe and $L_{6}$ for the conjugate, and a $2$-inch diameter lens with a focal length of 100~mm, labeled as $L_{7}$ for the probe and $L_{8}$ for the conjugate in Fig.~\ref{Fig:setup}.  The high NA lenses make it possible to capture as much of the light diffracting from the quadrant sensor and mask as possible.

It has been shown that a transmission-based sensing scheme with an optimized intensity-difference measurement, in which electronic gain or attenuation is used to compensate for the effect of optical losses, can reach the fundamental limit of sensitivity given by the quantum Cram{\'e}r-Rao bound~\cite{woodworth2020transmission,woodworth2022transmission}. In our experiment, given that the optical losses introduced by the plasmonic sensors are significantly larger than the ones introduced by the mask, electronic attenuation is used in the conjugate's photocurrent to minimize the intensity-difference noise between correlated quadrants (see Appendix).  We implement the optimized intensity-difference measurements with a  home-built quadrant detection system that provides access to the low (DC) and high (RF) frequency components of the photocurrent for each of the eight separate quadrants, four for the probe and four for the conjugate. This makes it possible to perform measurements with all possible combinations of quadrants. The quadrants of the photodetectors corresponding to the different quadrants of the plasmonic sensor and mask follow the same labeling convention, specifically $p_i$ and $c_j$ for the probe and conjugated photodetectors, respectively.  To characterize the response of the plasmonic sensors, the RF components of a pair of quadrant photocurrents, one from any of the probe's quadrants  $p_i$ and another one from the conjugate's quadrants  $c_j$, are subtracted using a hybrid junction and the resulting signal is measured with a spectrum analyzer. The DC components of the individual quadrants are used to monitor the optical power of the probing beams, and thus the sensing resources, and serve to verify the measured shot noise limit (see below).

\section{Results and Discussion}
Resonant plasmonic sensors experience a frequency shift of their EOT resonance in response to a local change in  refractive index.  This frequency shift leads to a corresponding transmission change for the probing field, which makes it possible to use them as sensors of the refractive index changes in their local environment~\cite{zhou2016performance,dowran2021fundamental}. Thus, in order to determine their sensitivity and the degree of quantum enhancement, we introduce a controlled modulation of the local index of refraction around each sensor. Considering the squeezing spectrum of our initial twin beams (see Fig.~\ref{Fig:bTMSS} in Appendix), we choose to perform our measurements at $400$~kHz, where the initial level of squeezing is maximum.  To do so, the nanohole array is placed inside a hermetically sealed chamber, as described in~\cite{dowran2018quantum}, where an ultrasonic buzzer (Multicomp: MCUSD-11A400B11RS, $400$~kHz resonance frequency) is used to generate pressure waves inside the chamber that in turn lead to local changes in the index of refraction of the air. Due to reflections of the ultrasound inside the chamber, a standing wave pattern is formed, which leads to a different modulation of the index of refraction at each of the four sensors. The index modulations are then transduced to intensity modulations on the probing field, such that the information from each sensor is accessible independently and simultaneously through the quadrant detection system. The magnitudes of these signals are proportional to the driving voltage of the ultrasonic buzzer as well as the response of individual plasmonic sensors.

\begin{figure}
  \centering
  \includegraphics{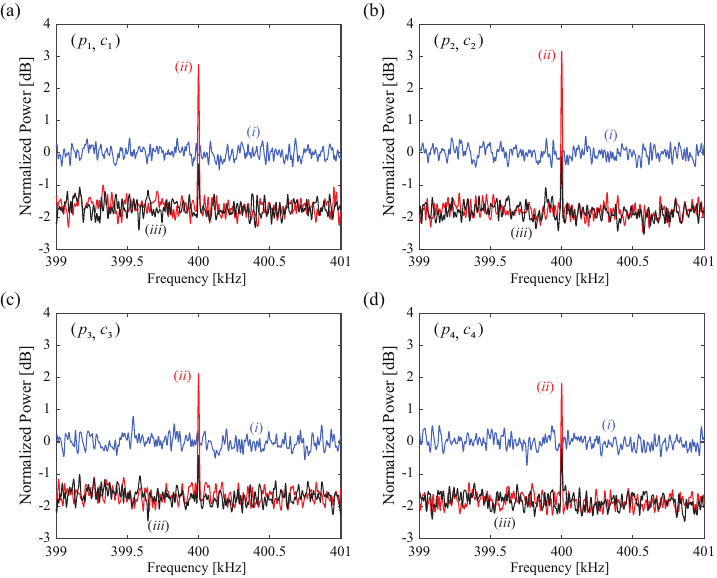}
  \caption{Parallel quantum-enhanced plasmonic sensing. An ultrasonic buzzer driven at its resonant frequency of $400$~kHz is used to introduce fixed amplitude modulations in the local refractive index of the air around each of the plasmonic sensors.  The peaks in the power spectra result from the transduction of index of refraction modulation to transmission modulation by the plasmonic sensors. The modulation amplitude of the buzzer's driving voltage is set to either $120$~mV, red traces labeled $(ii)$, or $60$~mV, black traces labeled $(iii)$. Measurements on correlated quadrant pairs ($i=j$) when probing with the twin beam lead to a reduction of the noise below the SNL, blue traces labeled $(i)$, with squeezing levels: $(p_1, c_1)=-1.69$~dB, $(p_2, c_2)=-1.81$~dB, $(p_3, c_3)=-1.7$~dB, and $(p_4, c_4)=-1.84$~dB. All traces are normalized to their corresponding SNL. Spectrum analyzer settings: resolution bandwidth (RBW): $10$~Hz, video bandwidth (VBW): $1$~Hz, power averaging: $200$~times.}
  \label{Fig:Discrete}
\end{figure}

We start by considering optimized intensity-difference measurements for photodetector quadrant pairs $(p_{i}, c_{j})$ that are expected to be correlated, that is, $i=j$. The resulting power spectra (normalized to the corresponding SNL)  are shown in Fig.~\ref{Fig:Discrete} for different modulation levels of the local index of refraction around the plasmonic array. Two different index of refraction modulation amplitudes are shown: trace $(ii)$ in red and trace ${(iii)}$ in black.  While the four plasmonic sensors consistently show a reduction of the signal as the modulation voltage applied to the ultrasonic buzzer decreases, the measured signal power is different for the same modulation voltage. As mentioned above, this is a result of the ultrasonic standing wave pattern that builds up inside the chamber and to a lesser degree to the slightly different response of each of the plasmonic sensors, shown in Fig.~\ref{Fig:PlasmonicSensor}(b).

As shown in Fig.~\ref{Fig:Discrete}, the presence of squeezing for all the quantum-correlated quadrants enhances the signal-to-noise ratio (SNR) and allows for the detection of weak modulation signals.  The level of quantum enhancement depends on the remaining level of squeezing after the plasmonic sensor array and mask for each quadrant.  As can be seen, the level of squeezing is measured to be $-1.69$~dB for quadrant pairs $(p_1, c_1)$, $-1.81$~dB for $(p_2, c_2)$, $-1.70$~dB for $(p_3, c_3)$, and $-1.84$~dB for $(p_4, c_4)$. While the source can directly produce $-5.16$~dB of squeezing, this level is reduced to $-4.75$~dB once the optical systems needed to focus into the plasmonic structures and then image to the quadrant photodiodes is put in place.  The level is further reduced to $-3.75$~dB when razor blades are used to select corresponding quadrants of the probe and conjugate beams at the plane of the quadrant plasmonic sensor. This last measurement shows the effect of having a finite-sized coherence area.  Finally, if we consider the probe transmission of $\sim50\%$ due to EOT through each quadrant of the plasmonic sensor, see Fig.~\ref{Fig:PlasmonicSensor}(b), and the $90\%$ transmission of the conjugate through each of the quadrants of the mask, we expect $\approx-1.92$~dB of squeezing to be left for each quadrant pair with an optimal electronic attenuation of $5.2$~dB (see Appendix).  This level of squeezing is consistent with the measured squeezing levels and experimentally used attenuation of $5.1$~dB to minimize the noise level. We have also verified that subtracting uncorrelated quadrant pairs, $(p_i, c_j)$ with $i\ne j$, leads to excess noise above the SNL, and that in this case the noise of the probe and conjugate quadrants adds in quadrature (see Appendix) as expected, given that those quadrants of the twin beams are uncorrelated. This result shows that the quadrants in the twin beam are spatially independent from each other.

For each quadrant pair, the SNL (blue traces labeled $(i)$ in Fig.~\ref{Fig:Discrete}) is measured by using coherent states with the same optical power as the portion of the probe and conjugate incident on the corresponding quadrants.  When determining the SNL, the optimum electronic attenuation for the conjugate detector is kept at the level needed to minimize the intensity-difference noise when probing with twin beams. Due to the imbalance introduced by the electronic attenuation, any classical excess noise above that of a pure coherent state in the beams used to perform the measurement can lead to a miscalibration of the SNL. While it is technically hard to produce pure coherent states of light, we can reach the SNL using a laser beam that does not undergo FWM. We verify that when performing the optimized intensity-difference measurements, with the gain optimized for probing with the twin beams, a seed probe beam that does not undergo FWM leads to a measured noise level that changes linearly with total optical power, as expected for a shot-noise limited measurement. Moreover, the measured SNL is within $0.2$~dB of the theoretical estimation of the SNL based on the measured DC values (see Appendix).

The results for discrete modulation signals shown in Fig.~\ref{Fig:Discrete} show that the twin beam can simultaneously enhance the sensitivity of the four plasmonic sensors to enable the detection of a modulation in the index of refraction below the SNL. Next, we quantify the quantum-based sensitivity enhancement by determining the smallest modulation in the index of refraction that can be resolved with twin beams as compared to coherent states. To do so, we linearly decrease the modulation amplitude of the driving voltage of the ultrasonic buzzer, which consequently decreases the amplitude of the modulation of the refractive index of air, and  determine the voltage at which the $\text{SNR}=1$. For these measurements, the spectrum analyzer is set to zero span at $400$~kHz and is synchronized with the function generator to record the decreasing modulation signal from each sensor.

\begin{figure}
  \centering
  \includegraphics{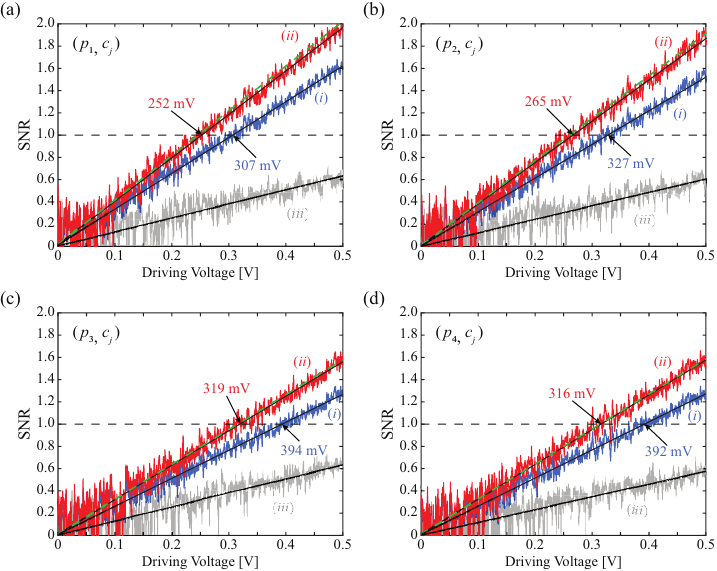}
  \caption{SNR as a function of modulation amplitude of the ultrasonic buzzer driving voltage. The minimum detectable change in the index of refraction is given by the voltage at which $\text{SNR}=1$ (horizontal dashed line).  The red (gray) traces show the SNR when the system is probed with twin beams for correlated (uncorrelated) quadrant pairs. The blue trace show the estimated SNR for the SNL for the equivalent classical configuration while the green dashed trace  provides the estimated SNR for the optimal classical configuration. Spectrum analyzer settings: measurement frequency: $400$~kHz, RBW: zero span, VBW: $10$~Hz, power averaging: $100$~times.}
  \label{Fig:SNR}
\end{figure}

To determine the SNR for each quadrant of the plasmonic array, we first estimated the modulation signal $S$ from the measured power spectrum by subtracting the measured noise power with no modulation, $s_{\text{off}}$, from the measured power with  modulation, $s_{\text{on}}$, that is
\begin{equation}
    S=s_{\text{on}}-s_{\text{off}}.\label{eq:signalTB}
\end{equation}
The SNR is then given by the square root of the ratio between the estimated signal to the measured noise power with no modulation,
\begin{equation}
    \text{SNR}=\sqrt{\frac{S}{s_{\text{off}}}}.\label{eq:SNLTB}
\end{equation}
We obtain the SNR for all 16 possible  quadrant pair combinations $(p_i, c_j)$ when probing with the twin beams as well as the SNR corresponding to the SNL. Figure~\ref{Fig:SNR} shows the SNR as a function of increasing modulation voltage amplitude for all cases, with the blue traces, labeled $(i)$, representing the SNL, the red traces, labeled $(ii)$, the correlated quadrants ($i=j$) when probing with twin beams, and the grey traces, labeled $(iii)$, the uncorrelated quadrants ($i\neq j$) when probing with twin beams. For clarity, each  subfigure in Fig.~\ref{Fig:SNR} only shows a single trace out of the four possible ones for the uncorrelated quadrant pairs. The SNRs for the other possible combinations show the same behavior as the one illustrated in the corresponding subfigure. As expected, only the four correlated quadrant combinations show a quantum enhanced SNR. Given the excess noise when measuring uncorrelated quadrants, the SNR is significantly degraded even with respect to the SNL when $i\neq j$.

While we estimate the SNR when probing with the twin beams directly from the measured power of the signal in the presence of a modulation, technical limitations prevent us from doing so for coherent states. In particular, this is due to the sensitivity of the measured signal level to the alignment and amount of optical power that couples to each quadrant in the plasmonic array. In order to measure the SNL we need to use an input probe beam that does not undergo the FWM process. As noted above, the probe beam experiences a cross-Kerr effect that changes its size at the sensing plane. Thus, the size of the probe beam changes in the absence of FWM, which leads to a different level of optical power coupling to each quadrant in the plasmonic array. While in principle it would be possible to modify the optical system to compensate for such a change, this would lead to a different alignment through the plasmonic structure and thus to a miscalibration of the SNL. We can, however, estimate the power of the modulation signal for probing with coherent states ($S^{\text{CS}}$) from the measurements performed with the twin beams ($S^{\text{TB}}$). In order to obtain an accurate estimation of the signal, as defined by Eq.~\eqref{eq:signalTB}, it is important to consider a regime in which the modulation is significantly larger than the noise. Under these conditions, we can assume that $S^{\text{TB}}\approx S^{\text{CS}}\equiv S$, such that for the coherent states
\begin{equation}
    \text{SNR}^{\text{CS}}\approx\sqrt{\frac{S}{s^{\text{CS}}_{\text{off}}}}, \label{eq:SNRCS}
\end{equation}
where $s^{\text{CS}}_{\text{off}}$ is the noise powered in the absence of a modulation when probing with coherent states. This leads to a more accurate estimation of the SNR for coherent states, and thus the SNL.

We estimate the degree of quantum enhancement for all four quadrants by first determining the minimum detectable change in the index of refraction given by the voltage at which the $\text{SNR}=1$ (horizontal black dashed line in Fig.~\ref{Fig:SNR}) when probing with coherent states and twin beams and then taking the ratio between them. As shown in Fig.~\ref{Fig:SNR}, for the case of correlated plasmonic and mask quadrants, these modulation voltages for the coherent states, $V_{\text{CS}}$, and the twin beams, $V_{\text{TB}}$, take values of $V_{\text{CS}}=307$~mV and $V_{\text{TB}}=252$~mV for $i=j=1$, $V_{\text{CS}}=327$~mV and $V_{\text{TB}}=265$~mV for $i=j=2$, $V_{\text{CS}}=394$~mV and $V_{\text{TB}}=319$~mV for $i=j=3$, and $V_{\text{CS}}=392$~mV and $V_{\text{TB}}=316$~mV for $i=j=4$.  These values show that a quantum enhancement in the range of 22\% to 24\% is obtained simultaneously for the four sensors when probed in parallel with the multi-spatial mode twin beams. These levels of quantum enhancement are consistent with the expected theoretical level given the remaining levels of squeezing~\cite{dowran2021fundamental} in the range from $-1.69$~dB to $-1.84$~dB. While these measurements do not provide an absolute measurement of the sensitivity to changes in the index of refraction, our previous studies have shown that the sensitivity of the plasmonic structures that we use is comparable to the state-of-the-art~\cite{dowran2018quantum}.

When performing quantum-enhanced measurements, it is also important to compare not only to the corresponding classical configuration, but also to the optimal classical measurement. As shown in~\cite{woodworth2020transmission}, the optimal classical measurement is one in which only a single beam coherent state is used to probe the system and no reference beam is used. This results from the fact that the reference beam carries no information and is not correlated to the probe beam, and as a result its presence only leads to additional noise in the measurement.  While in practice it is hard to perform such a measurement given that generating a pure coherent state is non-trivial, we estimate the SNR for the optimal classical configuration by using the same signal as the one used to calculate $\text{SNR}^{\text{CS}}$ above and adjusting the noise level by taking into account the reduction in total power due to the absence of a reference beam. The resulting SNRs for the optimal classical configuration are given by the dashed green traces in Fig.~\ref{Fig:SNR}. As we can see, even with the significant reduction in the level of squeezing due to the losses from the plasmonic structures and the cutting due to the quadrant configuration, we are able to perform at the level of the optimal classical measurement.  Thus, further enhancements in the fabrication of the structures to reduce the losses will easily make it possible to obtain sensitivities  beyond the optimal classical measurements.

\section{Conclusion}\label{conclusion}
We have shown that it is possible to take advantage of the temporal and spatial quantum properties of quantum states of light to implement a parallel quantum-enhanced sensing configuration. In particular, we use multi-spatial mode twin beams to probe a quadrant array of plasmonic sensors to independently and simultaneously detect local modulations in the refractive index of air below the SNL, thus implementing a spatially resolved quantum sensing configuration.  We show quantum enhancements in the range of 22\% to 24\% for all four sensors when probed in parallel with the twin beams. The levels of quantum enhancement are consistent with measured levels of squeezing of $- 1.69$~dB to $- 1.84$~dB.  These results  represent the first step toward applications that require spatially resolved quantum sensing, such as imaging and particle tracking, as well as fundamental studies of events that require complex arrays of quantum sensors, such as dark matter detection.

\section*{Acknowledgments}
This work was supported by the W. M. Keck Foundation and by a grant from the National Science Foundation  (NSF grant PHYS-1752938). The plasmonic sensor fabrication was supported in part by the U. S. Department of Energy, Office of Science, Basic Energy Sciences, Materials Sciences and Engineering Division. The electron beam lithography was supported by the Center for Nanophase Materials Sciences, which is a U.S. Department of Energy Office of Science User Facility.

\section*{Appendix}
\subsection*{Estimated Squeezing Level in the Presence of Losses with Optimized Intensity-Difference Measurements}
We estimate the expected level of squeezing remaining after probing a transmission-based sensor, such as the plasmonic sensor used in our experiments~\cite{woodworth2020transmission,dowran2021fundamental}, with twin beams.
Taking the transmissions of the probe and conjugate beams to be $\eta_p$ and $\eta_c$, respectively, the variance of the intensity-difference measurement can be shown to be given by
\begin{equation}
      \langle \Delta^2 I_{p-c}   \rangle^{\text{loss}} = \left[\eta_p^2 \left(  \langle \Delta^2 I_p   \rangle -  \langle I_p   \rangle \right)+ \eta_p \langle I_p \rangle \right]
    +g^2 \left[\eta_c^2  \left(  \langle \Delta^2 I_c   \rangle -  \langle I_c   \rangle  \right)+ \eta_c \langle I_c \rangle \right]
    -2G\eta_p\eta_c\text{Cov}, \label{eq:IDwithloss}
\end{equation}
where $\text{Cov}\equiv\langle I_p I_c \rangle - \langle I_p \rangle \langle I_c \rangle$ is the covariance and $g$ is an attenuation factor on the conjugate photocurrent. The optimal measurement that saturates the quantum Cram{\'e}r-Rao bound~\cite{woodworth2020transmission} is obtained when the attenuation $g$ is used to minimize this variance.

The optimal value of the electronic attenuation $g$ is obtained by minimizing Eq.~(\ref{eq:IDwithloss}) and can be shown to be given by
\begin{equation}
    g_{\text{opt}}=\frac{\eta_p\eta_c\text{Cov}}{ \eta_c^2  \left(  \langle \Delta^2 I_c   \rangle -  \langle I_c   \rangle  \right)+\eta_c \langle I_c \rangle }. \label{eq:optimumEA}
\end{equation}
With this optimal value, the minimum noise in the intensity-difference measurement takes the form
\begin{equation}
      \langle \Delta^2I_{p-c}   \rangle^{\text{loss}}_{\text{min}} = \left[\eta_p^2  \left(  \langle \Delta^2 I_p   \rangle -  \langle I_p   \rangle  \right)+\eta_p \langle I_p \rangle \right]
    - \frac{(\eta_p \eta_c \text{Cov})^2}{ \eta_c^2  \left(  \langle \Delta^2 I_c   \rangle -  \langle I_c   \rangle  \right)+\eta_c \langle I_c \rangle }. \label{eq:IDoptimum}
\end{equation}
Experimentally, the covariance term between the twin beams can be estimated by performing individual beam ($\langle \Delta^2 I_p  \rangle$ and $\langle \Delta^2 I_c\rangle$) and intensity-difference ($\langle \Delta^2 I_{p-c}  \rangle$) noise measurements with a balanced ($g=1$) homodyne detection configuration while bypassing the quadrant sensor and mask. Through the use of~\eqref{eq:IDwithloss} for the balanced $g=1$ case with no loss and the measured noise levels, the covariance of the twin beam can be calculated according to
\begin{equation}
      \text{Cov}= \frac{1}{2}\left( \langle \Delta^2 I_p  \rangle +   \langle \Delta^2 I_c   \rangle - \langle \Delta^2 I_{p-c}  \rangle \right). \label{eq:cov}
\end{equation}

\begin{figure}[h]
  \centering
  \includegraphics{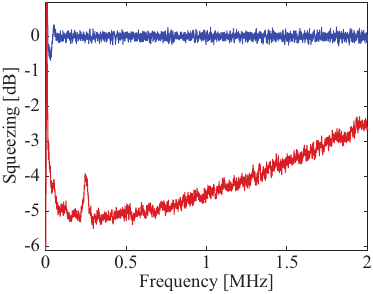}
  \caption{Squeezing spectrum of the FWM source (red trace) normalized to the SNL (blue trace). The FWM process was optimized to obtain independent quantum-correlated quadrants for the generated twin beams.}
  \label{Fig:bTMSS}
\end{figure}
The shot noise level (SNL) can be measured through the use of coherent states with the same optical power as the twin beams. To obtain a valid measure of the shot noise level the detection system has to be the same as the one used for the twin beams, which means that the electronic attenuation factor $g$ has to be set to the one that minimizes the noise measurement for the twin beam, as given by \eqref{eq:optimumEA}.  Given that the electronic attenuation factor leads to an imbalance in the detection system, excess noise in the optical beam used to calibrate the SNL beyond that of a coherent state can lead to a miscalibration. To minimize this effect, we use the probe beam without the FMW process.  To further validate the SNL that is measured, we compared the measured noise levels with the theoretically expected ones. To do so, we take into account the fact that there is no covariance between the two coherent states used for the SNL estimation and that their variance is equal to their mean values, $\langle \Delta^2 I\rangle= \langle I \rangle$, such that~\eqref{eq:IDwithloss} is simplified to
\begin{equation}
    \langle \Delta^2I_{p-c} \rangle^{\text{loss}}_{\text{CS}}= \eta_p \langle I_p \rangle_{\text{CS}}
    +g_{\text{opt}}^2\eta_c \langle I_c \rangle_{\text{CS}}.
\end{equation}
The noise-equivalent level for a given  optical power is calibrated through measurements of the noise power vs optical power that are then used to obtain the slope through a linear fit. Through this method, we find that the measured SNL matches well with the estimated values, with a difference of less than $0.2$ to $0.3$~dB.

The level of squeezing in linear scale, $R$, is then obtained by taking the ratio of the intensity-difference noise measurements with the twin beams to that one with coherent states, that is
\begin{equation}
   R=\frac{\langle \Delta^2I_{p-c}   \rangle^{\text{loss}}_{\text{min}}}{\langle \Delta^2I_{p-c} \rangle^{\text{loss}}_{\text{CS}}}.
\end{equation}
The initial level of squeezing generated by the source, with $g=1$ and $\eta_{p}=\eta_{c}=1$, is shown in Fig.~\ref{Fig:bTMSS}.  As can be seen, the source is able to generate $-5.16$~dB of squeezing, which is then degraded to range of $-1.7$~dB to $-1.8$~dB with an optimal value of $g=5.1$~dB, as described in the main manuscript.

\subsection*{Alignment Procedure: Location of Sensing and Detector Planes}
\begin{figure}[b]
  \centering
  \includegraphics{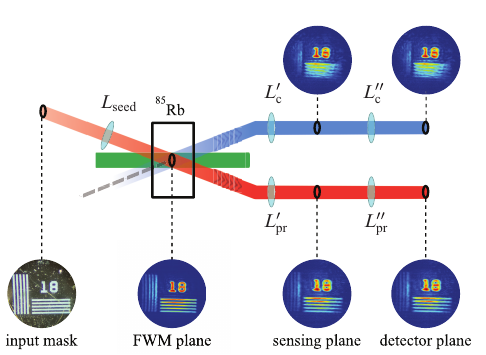}
  \caption{Alignment of optical systems for the sensing and detector planes. A mask with number $18$ is used to form images at the relevant experimental planes for proper alignment of the optical systems.}
  \label{Fig:alignment}
\end{figure}
Care has to be taken in determining the sensing and detector planes after the Rb cell due to a cross-Kerr effect between the strong pump and the twin beams that changes the location of the imaging plane.  Given that the probe is significantly closer to resonance than the pump, it experiences a stronger shift in the position of the imaging plane. To find the proper location of the imaging planes of the Rb cell for the probe and conjugate, we first image a target mask (Thorlabs NBS 1963A) with a pattern of number 18 at the center of the cell, as shown in Fig.~\ref{Fig:alignment}. Using the optimal parameters for the FWM, two separate imaging systems (one for the probe and one for the conjugate) are used to find the positions where the images are formed (see  Fig.~\ref{Fig:alignment}). These locations for the probe and conjugate correspond to the sensing plane, where the quadrant plasmonic sensors and mask are placed, respectively.  The focal lengths for the lenses used for these imaging systems are chosen to also produce the desired beam waists at the sensing plane when the probe has a Gaussian profile. We use the same procedure, with the mask in place, to properly align the optical systems for the probe and conjugate after the sensing plane to determine the location of detector plane. This ensures that the quadrant profile of the quadrant plasmonic structure and mask are properly detected by the quadrant photodiodes. In selecting the focal lengths for this last optical system, the required magnification was taken into account.

\subsection*{Independence of Spatial Quadrants of the Twin Beam}
\begin{figure}[h]
  \centering
  \includegraphics{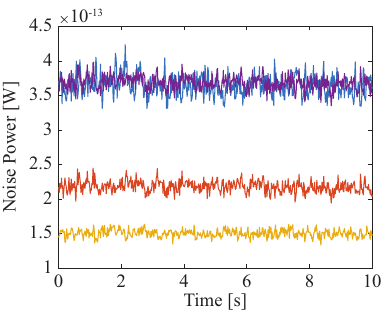}
  \caption{Noise power measurements to show independence of quadrants of twin beams.  The noise power of the intensity difference for uncorrelated pair $(p_3, c_4)$ (blue trace) is equal to the quadrature sum  (violet trace) of the noise powers of individual quadrants $p_3$ of the probe (yellow trace) and $c_4$ of the conjugate (red), which indicates the independence of these quadrants of the twin beam. Similar measurements for all uncorrelated quadrant pairs show they are spatially independent.}
  \label{Fig:QuadratureNoise}
\end{figure}

To verify that the four quadrants of the twin beams that interact with each of the quadrants of the plasmonic array are independent of each other, we perform noise measurements of two non-corresponding quadrants for the probe and conjugate.  In particular, if the non-corresponding quadrants of the twin beam are independent, then the noise of the intensity difference between them should be equal to the quadrature sum of the individual noise levels. We perform measurements of the different combinations of non-corresponding quadrants and find that this is the case, as shown in Fig.~\ref{Fig:QuadratureNoise} for quadrant pair $(p_3, c_4)$.

\subsection*{Optimal Transmission through Quadrant Array}
\begin{figure}[b]
  \centering
  \includegraphics{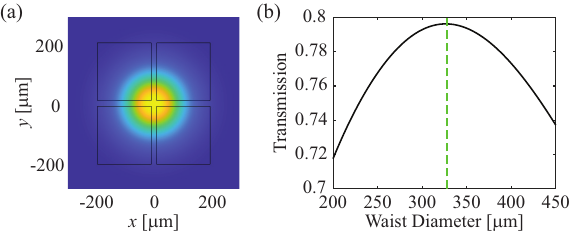}
  \caption{(a) Gaussian profile of optical beam incident on the quadrant sensor or mask. The quadrant window is tilted by $26^{\circ}$, consistent with the experimental implementation. (b) Optical transmission through quadrant configuration is maximized for a beam diameter waist of about $330\mu$m. }
  \label{fig:Gaussian}
\end{figure}
In order to determine the optimal size of the beam probing the quadrant sensor array such that losses due to the 20~$\mu$m gap between the plasmonic senors and the quadrant configuration are minimized, we consider a beam with a Gaussian profile
\begin{equation}
    \frac{1}{2\pi\sigma_x\sigma_y} e^{-\frac{x^2}{2\sigma_x^2}} e^{-\frac{y^2}{2\sigma_y^2}},
\end{equation}
where $\sigma_x$ and $\sigma_y$ are the beam sizes along Cartesian $x$ and $y$ coordinates, respectively, to calculate the expected transmission through a quadrant array with the dimensions of our sensor array and mask as a function of the beam size, as shown in Fig.~\ref{fig:Gaussian}(a). To better match the experiment, we consider  a  quadrant configuration that is tilted by $26^{\circ}$ about the $y$-axis, as implemented in the experiment to optimize the resonant frequency.
The resulting transmission, shown in Fig.~\ref{fig:Gaussian}(b), has a maximum of about $80\%$ for a beam with a $1/e^2$ waist diameter of about $330~\mu$m.


\begin{thebibliography}{10}
\newcommand{\enquote}[1]{``#1''}

\bibitem{giovannetti2004quantum}
V.~Giovannetti, S.~Lloyd, and L.~Maccone, \enquote{Quantum-enhanced
  measurements: beating the standard quantum limit,}
  {\protect\JournalTitle{Science}} \textbf{306}, 1330--1336 (2004).

\bibitem{lawrie2019quantum}
B.~J. Lawrie, P.~D. Lett, A.~M. Marino, and R.~C. Pooser, \enquote{Quantum
  sensing with squeezed light,} {\protect\JournalTitle{ACS Photonics}}
  \textbf{6}, 1307--1318 (2019).

\bibitem{dowling2003quantum}
J.~P. Dowling and G.~J. Milburn, \enquote{Quantum technology: the second
  quantum revolution,} {\protect\JournalTitle{Philosophical Transactions of the
  Royal Society of London. Series A: Mathematical, Physical and Engineering
  Sciences}} \textbf{361}, 1655--1674 (2003).

\bibitem{deutsch2020harnessing}
I.~H. Deutsch, \enquote{Harnessing the power of the second quantum revolution,}
  {\protect\JournalTitle{PRX Quantum}} \textbf{1}, 020101 (2020).

\bibitem{caves1981quantum}
C.~M. Caves, \enquote{Quantum-mechanical noise in an interferometer,}
  {\protect\JournalTitle{Physical Review D}} \textbf{23}, 1693 (1981).

\bibitem{slusher1985observation}
R.~Slusher, L.~Hollberg, B.~Yurke, J.~Mertz, and J.~Valley,
  \enquote{Observation of squeezed states generated by four-wave mixing in an
  optical cavity,} {\protect\JournalTitle{Physical Review Letters}}
  \textbf{55}, 2409 (1985).

\bibitem{pirandola2018advances}
S.~Pirandola, B.~R. Bardhan, T.~Gehring, C.~Weedbrook, and S.~Lloyd,
  \enquote{Advances in photonic quantum sensing,} {\protect\JournalTitle{Nature
  Photonics}} \textbf{12}, 724--733 (2018).

\bibitem{acsphotonics.5b00501}
R.~C. Pooser and B.~Lawrie, \enquote{Plasmonic trace sensing below the photon
  shot noise limit,} {\protect\JournalTitle{ACS Photonics}} \textbf{3}, 8--13
  (2016).

\bibitem{RN1331}
C.~Lee, B.~Lawrie, R.~Pooser, K.-G. Lee, C.~Rockstuhl, and M.~Tame,
  \enquote{Quantum plasmonic sensors,} {\protect\JournalTitle{Chemical
  Reviews}} \textbf{121}, 4743--4804 (2021).

\bibitem{lugiato2002quantum}
L.~A. Lugiato, A.~Gatti, and E.~Brambilla, \enquote{Quantum imaging,}
  {\protect\JournalTitle{Journal of Optics B: Quantum and Semiclassical
  Optics}} \textbf{4}, S176 (2002).

\bibitem{genovese2016real}
M.~Genovese, \enquote{Real applications of quantum imaging,}
  {\protect\JournalTitle{Journal of Optics}} \textbf{18}, 073002 (2016).

\bibitem{magana2019quantum}
O.~S. Maga{\~n}a-Loaiza and R.~W. Boyd, \enquote{Quantum imaging and
  information,} {\protect\JournalTitle{Reports on Progress in Physics}}
  \textbf{82}, 124401 (2019).

\bibitem{boyer2008entangled}
V.~Boyer, A.~M. Marino, R.~C. Pooser, and P.~D. Lett, \enquote{Entangled images
  from four-wave mixing,} {\protect\JournalTitle{Science}} \textbf{321},
  544--547 (2008).

\bibitem{brida2010experimental}
G.~Brida, M.~Genovese, and I.~R. Berchera, \enquote{Experimental realization of
  sub-shot-noise quantum imaging,} {\protect\JournalTitle{Nature Photonics}}
  \textbf{4}, 227--230 (2010).

\bibitem{taylor2013biological}
M.~A. Taylor, J.~Janousek, V.~Daria, J.~Knittel, B.~Hage, H.-A. Bachor, and
  W.~P. Bowen, \enquote{Biological measurement beyond the quantum limit,}
  {\protect\JournalTitle{Nature Photonics}} \textbf{7}, 229--233 (2013).

\bibitem{taylor2014subdiffraction}
M.~A. Taylor, J.~Janousek, V.~Daria, J.~Knittel, B.~Hage, H.-A. Bachor, and
  W.~P. Bowen, \enquote{Subdiffraction-limited quantum imaging within a living
  cell,} {\protect\JournalTitle{Physical Review X}} \textbf{4}, 011017 (2014).

\bibitem{treps2003quantum}
N.~Treps, N.~Grosse, W.~P. Bowen, C.~Fabre, H.-A. Bachor, and P.~K. Lam,
  \enquote{A quantum laser pointer,} {\protect\JournalTitle{Science}}
  \textbf{301}, 940--943 (2003).

\bibitem{spagnolo2012quantum}
N.~Spagnolo, L.~Aparo, C.~Vitelli, A.~Crespi, R.~Ramponi, R.~Osellame,
  P.~Mataloni, and F.~Sciarrino, \enquote{Quantum interferometry with
  three-dimensional geometry,} {\protect\JournalTitle{Scientific Reports}}
  \textbf{2}, 1--6 (2012).

\bibitem{vidrighin2014joint}
M.~D. Vidrighin, G.~Donati, M.~G. Genoni, X.-M. Jin, W.~S. Kolthammer, M.~Kim,
  A.~Datta, M.~Barbieri, and I.~A. Walmsley, \enquote{Joint estimation of phase
  and phase diffusion for quantum metrology,} {\protect\JournalTitle{Nature
  Communications}} \textbf{5}, 1--7 (2014).

\bibitem{lassen2007tools}
M.~Lassen, V.~Delaubert, J.~Janousek, K.~Wagner, H.-A. Bachor, P.~K. Lam,
  N.~Treps, P.~Buchhave, C.~Fabre, and C.~Harb, \enquote{Tools for multimode
  quantum information: modulation, detection, and spatial quantum
  correlations,} {\protect\JournalTitle{Physical Review Letters}} \textbf{98},
  083602 (2007).

\bibitem{baumgratz2016quantum}
T.~Baumgratz and A.~Datta, \enquote{Quantum enhanced estimation of a
  multidimensional field,} {\protect\JournalTitle{Physical Review Letters}}
  \textbf{116}, 030801 (2016).

\bibitem{muraviev2018massively}
A.~Muraviev, V.~Smolski, Z.~Loparo, and K.~Vodopyanov, \enquote{Massively
  parallel sensing of trace molecules and their isotopologues with broadband
  subharmonic mid-infrared frequency combs,} {\protect\JournalTitle{Nature
  Photonics}} \textbf{12}, 209--214 (2018).

\bibitem{im2009plasmonic}
H.~Im, A.~Lesuffleur, N.~C. Lindquist, and S.-H. Oh, \enquote{Plasmonic
  nanoholes in a multichannel microarray format for parallel kinetic assays and
  differential sensing,} {\protect\JournalTitle{Analytical Chemistry}}
  \textbf{81}, 2854--2859 (2009).

\bibitem{acimovic2014lspr}
S.~S. Acimovic, M.~A. Ortega, V.~Sanz, J.~Berthelot, J.~L. Garcia-Cordero,
  J.~Renger, S.~J. Maerkl, M.~P. Kreuzer, and R.~Quidant, \enquote{LSPR chip
  for parallel, rapid, and sensitive detection of cancer markers in serum,}
  {\protect\JournalTitle{Nano Letters}} \textbf{14}, 2636--2641 (2014).

\bibitem{thijssen2014parallel}
R.~Thijssen, T.~J. Kippenberg, A.~Polman, and E.~Verhagen, \enquote{Parallel
  transduction of nanomechanical motion using plasmonic resonators,}
  {\protect\JournalTitle{ACS Photonics}} \textbf{1}, 1181--1188 (2014).

\bibitem{taylor2016quantum}
M.~A. Taylor and W.~P. Bowen, \enquote{Quantum metrology and its application in
  biology,} {\protect\JournalTitle{Physics Reports}} \textbf{615}, 1--59
  (2016).

\bibitem{altenburg2018multi}
S.~Altenburg and S.~W{\"o}lk, \enquote{Multi-parameter estimation: global,
  local and sequential strategies,} {\protect\JournalTitle{Physica Scripta}}
  \textbf{94}, 014001 (2018).

\bibitem{proctor2018multiparameter}
T.~J. Proctor, P.~A. Knott, and J.~A. Dunningham, \enquote{Multiparameter
  estimation in networked quantum sensors,} {\protect\JournalTitle{Physical
  Review Letters}} \textbf{120}, 080501 (2018).

\bibitem{ashok2021IOP}
A.~Kumar, G.~Nirala, and A.~M. Marino,
  \enquote{Einstein{\textendash}{P}odolsky{\textendash}{R}osen paradox with
  position{\textendash}momentum entangled macroscopic twin beams,}
  {\protect\JournalTitle{Quantum Sci. Technol.}} \textbf{6}, 045016 (2021).

\bibitem{boyer2008generation}
V.~Boyer, A.~M.~Marino, and P.~Lett, \enquote{Generation of spatially broadband
  twin beams for quantum imaging,} {\protect\JournalTitle{Physical Review
  Letters}} \textbf{100}, 143601 (2008).

\bibitem{kumar2019spatial}
A.~Kumar and A.~Marino, \enquote{Spatial squeezing in bright twin beams
  generated with four-wave mixing: constraints on characterization with an
  electron-multiplying charge-coupled-device camera,}
  {\protect\JournalTitle{Physical Review A}} \textbf{100}, 063828 (2019).

\bibitem{dowran2018quantum}
M.~Dowran, A.~Kumar, B.~J. Lawrie, R.~C. Pooser, and A.~M. Marino,
  \enquote{Quantum-enhanced plasmonic sensing,} {\protect\JournalTitle{Optica}}
  \textbf{5}, 628--633 (2018).

\bibitem{PhysRevLett.110.156802}
B.~J. Lawrie, P.~G. Evans, and R.~C. Pooser, \enquote{Extraordinary optical
  transmission of multimode quantum correlations via localized surface
  plasmons,} {\protect\JournalTitle{Phys. Rev. Lett.}} \textbf{110}, 156802
  (2013).

\bibitem{holtfrerich2016toward}
M.~Holtfrerich, M.~Dowran, R.~Davidson, B.~Lawrie, R.~Pooser, and A.~Marino,
  \enquote{Toward quantum plasmonic networks,} {\protect\JournalTitle{Optica}}
  \textbf{3}, 985--988 (2016).

\bibitem{homola2003present}
J.~Homola, \enquote{Present and future of surface plasmon resonance
  biosensors,} {\protect\JournalTitle{Analytical and Bioanalytical Chemistry}}
  \textbf{377}, 528--539 (2003).

\bibitem{stewart2008nanostructured}
M.~E. Stewart, C.~R. Anderton, L.~B. Thompson, J.~Maria, S.~K. Gray, J.~A.
  Rogers, and R.~G. Nuzzo, \enquote{Nanostructured plasmonic sensors,}
  {\protect\JournalTitle{Chemical Reviews}} \textbf{108}, 494--521 (2008).

\bibitem{mejia2018plasmonic}
J.~Mej{\'\i}a-Salazar and O.~N. Oliveira~Jr, \enquote{Plasmonic biosensing:
  focus review,} {\protect\JournalTitle{Chemical Reviews}} \textbf{118},
  10617--10625 (2018).

\bibitem{zoller2005quantum}
P.~Zoller, T.~Beth, D.~Binosi, R.~Blatt, H.~Briegel, D.~Bruss, T.~Calarco,
  J.~I. Cirac, D.~Deutsch, J.~Eisert \emph{et~al.}, \enquote{Quantum
  information processing and communication,} {\protect\JournalTitle{The
  European Physical Journal D-Atomic, Molecular, Optical and Plasma Physics}}
  \textbf{36}, 203--228 (2005).

\bibitem{PhysRevD.102.072003}
D.~Carney, S.~Ghosh, G.~Krnjaic, and J.~M. Taylor, \enquote{Proposal for
  gravitational direct detection of dark matter,} {\protect\JournalTitle{Phys.
  Rev. D}} \textbf{102}, 072003 (2020).

\bibitem{thewindchimecollaboration2022snowmass}
T.~W. Collaboration, A.~Attanasio, S.~A. Bhave, C.~Blanco, D.~Carney,
  M.~Demarteau, B.~Elshimy, M.~Febbraro, M.~A. Feldman, S.~Ghosh, A.~Hickin,
  S.~Hong, R.~F. Lang, B.~Lawrie, S.~Li, Z.~Liu, J.~P.~A. Maldonado,
  C.~Marvinney, H.~Z.~Y. Oo, Y.-Y. Pai, R.~Pooser, J.~Qin, T.~J. Sparmann,
  J.~M. Taylor, H.~Tian, and C.~Tunnell, \enquote{Snowmass 2021 white paper:
  The windchime project,} {\protect\JournalTitle{arXiv:2203.07242}}  (2022).

\bibitem{wu2021two}
M.-C. Wu, N.~R. Brewer, R.~W. Speirs, K.~M. Jones, and P.~D. Lett,
  \enquote{Two-beam coupling in the production of quantum correlated images by
  four-wave mixing,} {\protect\JournalTitle{Optics Express}} \textbf{29},
  16665--16675 (2021).

\bibitem{ebbesen1998extraordinary}
T.~W. Ebbesen, H.~J. Lezec, H.~Ghaemi, T.~Thio, and P.~A. Wolff,
  \enquote{Extraordinary optical transmission through sub-wavelength hole
  arrays,} {\protect\JournalTitle{Nature}} \textbf{391}, 667--669 (1998).

\bibitem{mccormick2008strong}
C.~McCormick, A.~M. Marino, V.~Boyer, and P.~D. Lett, \enquote{Strong
  low-frequency quantum correlations from a four-wave-mixing amplifier,}
  {\protect\JournalTitle{Physical Review A}} \textbf{78}, 043816 (2008).

\bibitem{hemmer1995efficient}
P.~Hemmer, D.~Katz, J.~Donoghue, M.~Cronin-Golomb, M.~Shahriar, and P.~Kumar,
  \enquote{Efficient low-intensity optical phase conjugation based on coherent
  population trapping in sodium,} {\protect\JournalTitle{Optics Letters}}
  \textbf{20}, 982--984 (1995).

\bibitem{holtfrerich2016control}
M.~Holtfrerich and A.~Marino, \enquote{Control of the size of the coherence
  area in entangled twin beams,} {\protect\JournalTitle{Physical Review A}}
  \textbf{93}, 063821 (2016).

\bibitem{martinelli2003experimental}
M.~Martinelli, N.~Treps, S.~Ducci, S.~Gigan, A.~Ma{\^\i}tre, and C.~Fabre,
  \enquote{Experimental study of the spatial distribution of quantum
  correlations in a confocal optical parametric oscillator,}
  {\protect\JournalTitle{Physical Review A}} \textbf{67}, 023808 (2003).

\bibitem{woodworth2020transmission}
T.~S. Woodworth, K.~W.~C. Chan, C.~Hermann-Avigliano, and A.~M. Marino,
  \enquote{Transmission estimation at the Cram{\'e}r-Rao bound for squeezed
  states of light in the presence of loss and imperfect detection,}
  {\protect\JournalTitle{Physical Review A}} \textbf{102}, 052603 (2020).

\bibitem{woodworth2022transmission}
T.~S. Woodworth, C.~Hermann-Avigliano, K.~W.~C. Chan, and A.~M. Marino,
  \enquote{Transmission estimation at the quantum Cram{\'e}r-Rao bound with
  macroscopic quantum light,} {\protect\JournalTitle{EPJ Quantum Technol.}}
  \textbf{9}, 38 (2022).

\bibitem{zhou2016performance}
X.~Zhou, L.~Zhang, and W.~Pang, \enquote{Performance and noise analysis of
  optical microresonator-based biochemical sensors using intensity detection,}
  {\protect\JournalTitle{Optics Express}} \textbf{24}, 18197--18208 (2016).

\bibitem{dowran2021fundamental}
M.~Dowran, T.~S. Woodworth, A.~Kumar, and A.~Marino, \enquote{Fundamental
  sensitivity bounds for quantum enhanced optical resonance sensors based on
  transmission and phase estimation,} {\protect\JournalTitle{Quantum Science
  and Technology}} \textbf{7}, 015011 (2022).

\end{thebibliography}
\end{document}